\documentclass[aps,prd]{revtex4}

\usepackage{amsmath, amssymb, bm, graphicx, graphics, color,mathrsfs,hyperref}

\newcommand{\be}{\begin{equation}}
\newcommand{\ee}{\end{equation}}
\newcommand{\ben}{\begin{eqnarray}}
\newcommand{\een}{\end{eqnarray}}

\newcommand{\cO}{{\cal O}}

\newcommand{\cH}{{\cal H}}
\newcommand{\cE}{{\cal E}}

\newcommand{\cB}{{\cal B}}
\newcommand{\cJ}{{\cal J}}

\newcommand{\p}{\partial}
\newcommand{\na}{\nabla}

\newcommand{\ep}{\epsilon}

\newcommand{\talpha}{{\tilde \alpha}}

\begin{document}

\title{Conductivity bound of the strongly interacting and disordered graphene from gauge/gravity duality}

\author{Marek Rogatko}
\email{marek.rogatko@poczta.umcs.lublin.pl,
rogat@kft.umcs.lublin.pl}
\author{Karol I. Wysoki\'nski}
\email{karol.wysokinski@poczta.umcs.pl}
\affiliation{Institute of Physics \protect \\
Maria Curie-Sklodowska University \protect \\
20-031 Lublin, pl.~Marii Curie-Sklodowskiej 1, Poland}

\date{\today}

\begin{abstract}
The carriers in graphene tuned close to the Dirac point 
envisage signatures of the strongly interacting fluid and
are subject to hydrodynamic description. The important question is whether strong disorder 
induces the metal - insulator transition in this two-dimensional material. 
The  bound on the conductivity tensor found 
earlier within the single current description, implies that the system does not 
feature metal - insulator transition. The linear
spectrum of the graphene imposes the phase - space constraints and calls 
for the two - current description of interacting electron and hole liquids. 
Based on the gauge/gravity correspondence, using the linear response of the black brane with broken
translation symmetry in Einstein-Maxwell gravity with the auxiliary $U(1)$-gauge field,
responsible for the second current, we have calculated the lower bound 
of the DC-conductivity in holographic model
of graphene. The calculations show that the bound on the conductivity depends 
on the coupling between both $U(1)$ fields and for a physically justified range
of parameters it departs only weakly from the value found for a model with the single $U(1)$ field.
\end{abstract}

\maketitle

\section{Introduction}
\label{sec:intro}

Disorder and interactions inside solids are responsible for finite values of the transport coefficients and
play a very important role in establishing their detailed behavior. Importantly, the role of both disorder 
and interactions  depends on  the spatial dimensionality 
of the condensed matter system.  Doping of the intrinsic semiconductors,
being the key ingredient of numerous electronic applications, is an important example
illustrating the role of disorder in the weakly interacting  three-dimensional materials. 

It has been predicted and verified experimentally that in three-dimensional systems both strong disorder or strong electron-electron
interactions can induce metal to insulator transition.  In the non-interacting systems 
this phenomenon is called the Anderson  transition \cite{anderson1958}, 
while in the presence of electron-electron interaction, the transition is known 
as Mott \cite{mott1968} or if interactions and disorder play a significant role, 
the Anderson-Hubbard one \cite{belitz1994}. 

On the other hand, the two-dimensional systems are far more complicated from the experimental  
\cite{kravchenko1994,melnikov2019,shashkin2019}, as well as, the theoretical points of view.  
The theoretical description of the 
interacting \cite{kravchenko2004} systems in question, does not give unique results.  
The recent application of the gauge/gravity analogy to study the strongly interacting  
two-dimensional disordered materials has revealed the absence of the disorder driven 
metal-insulator transition in the system \cite{gro15}. The result is valid in the hydrodynamic
limit for the electron mean free path much smaller than the typical scale of the 
spatial inhomogeneities \cite{lucas17}.

The hydrodynamic limit of the electron flow has been identified experimentally in very clean systems,
as predicted long time ago \cite{gurzhi1968}.
In fact, the signatures of the hydrodynamic behavior have been observed over the last years in many
materials including the high mobility (Al,Ga)As wires \cite{molenkamp1994,dejong1995}. More recent measurements 
have envisaged the hydrodynamic signatures in many other materials. One should mention the shear viscosity 
measurements in the  ultra-cold Fermi gases \cite{cao11}, strongly correlated  oxides \cite{moll2016} and  
graphene~\cite{crossno2016,bandurin2016}. The comprehensive discussion of this novel  set of experiments 
is given in~\cite{zaanen2016}. 

The special interest is devoted to graphene, the two-dimensional system which envisages 
a hydrodynamic behavior of the carriers, observed in a number of recent experiments
\cite{ghahari2016,krishna2017,bandurin2018,mayzel2019}, especially  when the material 
is tuned close to the particle-hole symmetry point. Due to the strong scattering of charge 
carriers, in the nearby of the charge neutrality point, the thermoelectric power 
of graphene is strongly enhanced \cite{ghahari2016} and 
approaches the hydrodynamic limit.  The  departures
from the Wiedemann-Franz ratio due to the increase of thermal 
conductivity \cite{crossno2016}, have been interpreted as the indication of hydrodynamic behavior 
in material in question.  On the other hand, the hydrodynamic viscosity
of electrons has been measured \cite{bandurin2016} in a high mobility graphene samples. 
The viscous effects were observed \cite{krishna2017} and   
shown to facilitate high mobility transport at temperatures below 150K. The recent theoretical 
and experimental studies of hydrodynamic effect in graphene have 
been reviewed in \cite{narozhny2017,lucas2018}. 

Even though the  hydrodynamic flow is expected to be observed in a very clean system, the 
disorder seems to be an important factor which sometimes even facilitates the hydrodynamic 
behavior \cite{zarenia2019}. 
The signatures of the Stokes non-linear flow with the   
low Reynolds number \cite{rey}
have been detected in graphene \cite{mayzel2019}, as the appearance 
of vortices leading to the negative resistance of the material.

At the Fermi level graphene exhibits a massless relativistic spectrum with Dirac cone. As was mentioned above, 
close to the charge neutrality point, it sustains a
strongly interacting material, ideal system for studies by means of gauge/gravity duality methods.  
In this system, the thermoelectric transport coefficients
have been found using the hydrodynamic approach \cite{muller2008,muller2009,lucas2016}, with a fairly 
good agreement with the experimental data.

Recently, this attitude has been generalized to the model with two distinct $U(1)$-gauge currents, 
which is solved by the AdS/CFT analogy \cite{seo17}.
The model in question allowed the successful quantitative comparison between theory and experimental data.
The paper \cite{seo17} gives a number of arguments behind the introduction of two gauge fields and associated currents.
One reason for the appearance of two currents in graphene is the charge imbalance between electrons and holes
in the system with linear spectrum. It has been found that the two current model allows for a quantitatively 
correct description of the thermal conductance of  graphene.  The paper \cite{rog18} presented the further generalization, taking into
account the possible coupling between both currents.
In Ref.\cite{rog18} the transport properties of graphene using the model in question were elaborated. Moreover the perpendicular magnetic field to the graphene 
sheet was taken into account. It was assumed that the charges bounded with the two gauge fields are proportional to each other with the factor $g$, 
which is responsible for the possible imbalance of the positive and negative charges in graphene close to the charge neutrality point. It was found that the kinetic and transport coefficients were influenced by $\alpha$-coupling constant and factor $g$. The increase of $\alpha$ leads to the increase of the width of normalized thermal conductivity, while in the case when $g=0$, the effect has been quite opposite (we have the decrease of the width). On the other hand, the $\alpha$-coupling constant affects Wiedeman-Franz ratio (WFR), changing the width and heights of the curves. The general tendency is that WFR diminishes while the value of $\alpha$-coupling constant grows.

Moreover the coupling constant in question, impels the charge dependence of the diagonal resistivity and the WFR, i.e., the increase of $\alpha$ causes the decrease of both $\rho^{xx}$ and $W^{xx}$. The Seebeck and Nernst coefficients were affected by magnetic field and $\alpha$. The influence in question, for large value of $S^{xx}$, changes the shape of the curve from two minima and a maximum curve to the one with a minimum (for $B=0$) and two small maxima for larger absolute values.
The Hall angle was also influenced by the coupling constant. In the studied case the density dependence of the thermoelectric coefficient $\alpha_{ij}$ and Seebeck coefficient $S^{xx}$ agree with experimental data.

The generalizations of these researches were given in \cite{rog18a}, where the holographic calculation of magneto transport coefficients in $3+1$ dimensional system with Dirac-like spectrum were presented. The calculations envisage the influence of $g$ and $\alpha$ on the coefficients. Namely, the magnetic field dependence on resistivity
$\rho^{xx}$ and $\rho^{xy}$ depicts that the bigger values of $\alpha$ one takes the smaller resistivity we achieve.

In general one expects the presence of additional gauge fields 
in graphene due to geometric and other reasons \cite{vozmediano2010,volovik2014}.   The use of gauge/gravity 
duality allows for the exact solution of the strongly coupled field theoretical models. 
We use this approach to elaborate
the effect of interactions and disorder on the hydrodynamic transport of graphene modeled by the 3+1 dimensional
AdS space time with the black brane background which breaks translational symmetry \cite{don15,ban15}.

The studies of electrical transport in a strongly coupled system include the case of strange metals in two spatial 
dimensions at finite temperature and charge density,
holographically dual to Einstein-Maxwell theory with a potential in asymptotically four-dimensional 
AdS manifold. One finds that the electrical conductivity is bounded
from below by a universal minimum conductance. The inspection of Stokes-like equations 
in the spacetime in question shows that it cannot exhibit 
metal-insulator transitions \cite{gro15}.
The bound on the incoherent thermal conductivity obtained by analyzing the linear perturbations 
of black brane with broken translation symmetry in AdS Einstein-Maxwell dilaton
gravity was performed in \cite{gro16}.  It turns out that the thermal conductivity 
has non-zero value (at finite temperature), as far as, the dilaton potential is bounded from below.

In \cite{fad16} the analytical lower bound on the conductivity in holographic model 
AdS Einstein-Maxwell dilaton theory, in terms of black horizon data, using the Stokes 
equations on black object event horizon was provided. In the considered model the metal-insulator 
transition is not driven by disorder, but it is caused by coupling scalar field to Maxwell one. Studies 
in rotational and translational symmetries breaking system reveal that the ratio of the determinant 
of the electrical conductivities along any spatial directions, to black brane area density, having the zero 
charge limit in account, tends to the universal value \cite{ge17}. The conductivity bounds were also elaborated 
in the case of probe brane models \cite{ike16}, massive gravity \cite{bag17}, as well as, in effective 
holographic theories \cite{gou16}. It was shown in the two latter
cases that there were no bounds on conductivities.

In \cite{rog19}  the Navier-Stokes equations of the model with
two $U(1)$-gauge fields were derived. The paper elaborates  the black brane response to the electric 
fields and temperature gradient. The DC transport coefficients for the holographic Dirac semimetals are found.
Here we analyze similar model with a goal to establish the bounds on the conductivity of the Dirac fluid in graphene
subject to the influence of $\alpha$-coupling constant between the two $U(1)$-gauge fields. The main result is that the coupling 
between the currents, in the following quantified by the
parameter $\alpha$ only slightly modifies the bounds (see Eqs. (\ref{bound1}) and (\ref{bound2})) on the conductivity 
for $\alpha \le 1$. Larger 
values of $\alpha$ lead to strong decrease of the bound and finally to metal - insulator transition
at $\alpha=2$, when the conductivity bound vanishes.

The organization of the paper is as follows. In Sec. \ref{sec:model} we introduce the gravitational
background and the action used to describe two interacting  currents in graphene. Sec. \ref{sec:pert}
is devoted to the description of the perturbations of the event horizon allowing the derivation of the
appropriate hydrodynamic description. We calculate the conductivity of the system in the background of the 
uncharged black brane in Sec. \ref{sec:cond}. In section \ref{sec:charged} the case of the charged 
background black brane is discussed, where we also derive the lower bounds on the conductivity.
The variational approach has been applied in Sec. \ref{sec:variation} to study the conductivity bounds. In 
Sec. \ref{sum-concl} we end up with the summary and conclusions. 

\section{Background holographic model}\label{sec:model}
In our paper we deal with the generalization of the previously studied models \cite{don15,ban15}, 
by adding two interacting $U(1)$-gauge fields. The aim is to
find the influence of them on DC thermoelectric transport coefficients and to compare with the existing results.
In our model the gravitational action in $(3+1)$-dimensions is taken in the form  
\be
S = \int \sqrt{-g}~ d^4 x~  \bigg( R + \frac{6}{L^2} - \frac{1}{2} \na_\mu \phi \na^\mu \phi
- \frac{1}{4} F_{\mu \nu} F^{\mu \nu} - \frac{1}{4}B_{\mu \nu} B^{\mu \nu} - \frac{\alpha}{4} F_{\mu \nu} B^{\mu \nu} \bigg),
\label{sgrav} 
\ee
where $R$ is the scalar curvature of the spacetime,
$\phi$ stands for the scalar field, which as we shall see later on contributes a viscosity like term 
to the hydrodynamic equations. $F_{\mu \nu} = 2 \nabla_{[ \mu} A_{\nu ]}$ are ordinary 
Maxwell field strength tensor, while
the second $U(1)$-gauge field $B_{\mu \nu}$ is given by $B_{\mu \nu} = 2 \nabla_{[ \mu} B_{\nu ]}$. 
$\alpha$ is the coupling constant between both gauge fields. $L$ is the radius of AdS-spacetime. 

The presence of additional gauge field 
is motivated by the desire of describing carrier flow in graphene, near the particle-hole symmetry point.
These two currents may be interpreted as connected with electrons and holes.
 The approach in question provides quantitatively correct description of the thermal conductivity 
of graphene close to the Dirac point \cite{seo17}. Allowing 
 for the interaction between the two $U(1)$-gauge currents, the coupling $\alpha$  provides additional 
degree of freedom and {\it inter alia} affects \cite{rog18} 
 the magnetic field dependence of the non-diagonal transport coefficients, especially 
for the low values of the aforementioned field.
The important novel aspect of the two current model is the tensor structure \cite{seo17,rog18} of the transport coefficients with the general entries, e.g., for the conductivity $\sigma_{ab}^{ij}$, where $a,b$ refer to two fields denoted above  as $F$ and $B$ and $i,j$ refer to the spatial directions (c.f. Eqs. (\ref{curf}) and (\ref{curb})). The  identifications of the charges $Q_F=-en_e$ and $Q_B=+en_h$ with electrons and holes and the total electric current  $J^j=J^j_F+J^j_B$ as well as assuming that the electric fields $E_F^i=E_B^i=E^i$ lead to the value of the total conductivity elements $\sigma^{ij}=\sum_{a,b}\sigma^{ij}_{ab}$. The presence of the  coupling $\alpha$ between the fields leads to non-zero values of $\sigma_{FB}^{ij}$. Moreover, independently whether the coupling  vanishes or not it is important to keep the tensor structure of the kinetic and transport coefficients \cite{seo17,rog18}. The  analogous studies of the magneto-transport coefficients of  Dirac semimetals \cite{rog18a}, being the three-dimensional analogues of graphene require similar treatment of the conductivity.  In both cases, in order to define the other transport coefficients, like thermoelectric tensor or Hall angle, one needs to take the full tensorial character of the conductivity into consideration.

In the studied action (\ref{sgrav}) we have to do with the second gauge field coupled to the ordinary Maxwell one. 
The justifications of such kind of gravity with electromagnetism coupled to the other gauge field 
follow from the top-down perspective \cite{ach16}. Namely, starting from
the string/M-theory the reduction to the lower dimensional gravity is performed. 
It is relevant in the holographic correspondence attitude, because the theory 
in question is a fully consistent quantum theory (string/M-theory) and this fact 
 guarantees  that any predicted phenomenon by the top-down theory will be physical. This point has been discussed in \cite{rog19}.

Variation of the action $S$ with respect
to the metric, the scalar and gauge fields yields the following equations of motion:
\ben 
G_{\mu \nu} &-& g_{\mu \nu}~\frac{3}{L^2}  = T_{\mu \nu}(\phi) + T_{\mu \nu}(F) + T_{\mu \nu}(B) 
+ \alpha~T_{\mu \nu}(F,~B),\\ \label{ff1}
\na_{\mu}F^{\mu \nu} &+& \frac{\alpha}{2}~\na_\mu B^{\mu \nu} = 0,\\ \label{bb1}
\na_{\mu}B^{\mu \nu} &+& \frac{\alpha}{2}~\na_\mu F^{\mu \nu} = 0,\\
\na_\mu \na^\mu \phi &=& 0,
\een
where we have denoted by $G_{\mu \nu}$ the Einstein tensor, while the energy momentum tensors for the fields in the theory are given by
\ben
T_{\mu \nu} (\phi) &=& \frac{1}{2} \na_{\mu} \phi \na_\nu \phi - \frac{1}{4}~g_{\mu \nu}~\na_\delta \phi \na^\delta \phi ,\\
T_{\mu \nu}(F) &=& \frac{1}{2}~F_{\mu \delta}F_{\nu}{}^{\delta} - \frac{1}{8}~g_{\mu \nu}~F_{\alpha \beta}F^{\alpha \beta},\\
T_{\mu \nu}(B) &=& \frac{1}{2}~B_{\mu \delta}B_{\nu}{}^{\delta} - \frac{1}{8}~g_{\mu \nu}~B_{\alpha \beta}B^{\alpha \beta},\\
T_{\mu \nu}(F,~B) &=& \frac{1}{2}~F_{\mu \delta}B_{\nu}{}^{\delta} - \frac{1}{8}~g_{\mu \nu}~F_{\alpha \beta}B^{\alpha \beta}.
\een

For the gauge fields in the considered theory we assume the following  components:
\be
A_\mu ~dx^\mu = a_t ~dt, \qquad
B_\mu~dx^\mu  = b_t~dt.
\ee

\section{Perturbations of background black brane}
\label{sec:pert}
In the following  analysis  we  consider the  line element provided by
\be
ds^2 = -U(r) G(r,x_i)dt^2 + \frac{F(r,x_i)dr^2}{U(r)} + ds^2(\Sigma_2),
\ee
where $\Sigma_2$ stands for the two-dimensional hypersurface at chosen radial $r$-coordinate. 
The dependence of function $G$  and $F$ on the $x_i$ coordinates takes care of their spatial 
variations. We also take the following  components of the fields
\be
A=a_t~dt, \quad  \quad B=b_t~dt.
\ee
As in \cite{don15}, the line element at $r \rightarrow \infty$ approaches the AdS boundary with the following conditions:
\ben
U \rightarrow r^2, \qquad F \rightarrow 1, \qquad G \rightarrow G(x), \qquad g_{ij} \rightarrow r^2~{\bar g}_{ij}\\
a_t(r, x_i) \rightarrow \mu(x), \qquad b_t(r, x_i) \rightarrow \mu_d(x), \qquad \phi(r, x_i) \rightarrow r^{\Delta -3} 
{\bar \phi}(x_i),
\een
where $\mu(x)$ and $\mu_d(x)$ are the spatially dependent chemical potentials (at the boundary) 
connected with the adequate $U(1)$-gauge field. We also assume  the periodic boundary conditions 
with period $L_i$ in the i-th direction: $f(x_i+L_i)=f(x_i)$ and if required shall work with quantities
averaged over the volume of periodicity $ \mathbb{E}[f]=\frac{1}{L_{x_1} L_{x_2}} \int dx_1dx_2 f$.
${\bar \phi}(x_i)$ above serves as a boundary source of the field $\phi(r, x_i)$ and $\Delta$ 
is the scaling dimension of it. 

The black brane event horizon which has $\Sigma_2$ topology, is situated at $r=0$. 
Having in mind the Edington-Finkelstein ingoing coordinates, the near-horizon
expansions of the metric tensor components and fields are given by \cite{ban15}
\ben
U(r) &=& r \Big( 4 \pi T + U^{(1)} r + \dots \Big),\label{hor-U}\\
G(r,x_i) &=& G^{(0)}(x) + G^{(1)}(x) r + \dots, \\
F(r, x_i) &=& F^{(0)}(x) + F^{(1)}(x) r + \dots, \\
g_{ij} &=& g^{(0)}_{ij} + g^{(1)}_{ij} r + \dots,\\
a_t(r, x_i) &=& r \Big( a^{(0)}_t ~G^{(0)}(x) + a^{(1)}_t(x) r + \dots \Big),\\
b_t(r, x_i) &=& r \Big( b^{(0)}_t ~G^{(0)}(x) + b^{(1)}_t(x) r + \dots \Big),\\
\phi(r, x_i) &=& \phi^{(0)}(x) + \phi^{(1)} (x) r + \dots,
\label{hor-exp}
\een
with the auxiliary condition written as $G^{(0)}(x) = F^{(0)}(x)$.

If one implements the $U(1)$-gauge and temperature gradient in the black brane spacetime, 
at fixed $r$-coordinate, then the black object will respond.  
In our considerations we have to take into account linear perturbations described by \cite{ban15}
\ben
\delta \Big(ds^2 \Big)  &=& \delta g_{\alpha \beta}~ dx^\alpha dx^\beta - 2 t~M~\xi_a dt~dx^a,\\
\delta A &=& \delta a_\beta ~dx^\beta - t~E_a dx^a + t~N~\xi_b~ dx^b,\\
\delta B &=& \delta b_\beta~ dx^\beta - t~B_a dx^a + t~N_d~\xi _b ~dx^b,
\een
as well as the perturbation of scalar field, $\delta \phi$. In what follows we shall consider 
$\delta g_{\mu \nu},~\delta a_\mu,~\delta b_\mu$, and $\delta \phi$ as 
functions of $(r,~x_m)$. On the other hand $E_a,~B_a, ~\xi_i$, depend on $x_i$-coordinates and are closed forms on $\Sigma_2$.
Moreover, the regularity at the black brane event horizon implies the following:
\ben \label{hor-pert-gtt}
\delta g_{tt} &=& U(r) \Big( \delta g^{(0)}_{tt}(x_i) + \cO(r) \Big), \qquad  \delta g_{tr} = \delta g^{(0)}_{tr }(x_i) + \cO(r),\\
\delta g_{rr} &=& \frac{1}{U(r)} \Big( \delta g^{(0)}_{rr}(x_i) + \cO(r) \Big), \qquad \delta g_{ij} = \delta g^{(0)}_{ij}(x_i) + \cO(r),\\
\delta g_{ti} &=& \delta g^{(0)}_{ti}(x_i) - GU \xi_i \frac{\ln r}{4 \pi T} + \cO(r), \qquad \delta g_{ri} = \frac{1}{U(r)} \Big( \delta g^{(0)}_{ri} (x_i) + \cO(r) \Big),\\
\delta a_t &=& \delta a^{(0)}_t (x_i)+ \cO(r), \qquad \delta a_i = \frac{\ln r }{4 \pi T} \Big( - E_i + N \xi_i \Big) + \cO(r),\\
\delta a_r &=& \frac{1}{U(r)} \Big( \delta a^{(0)}_r(x_i) + \cO(r) \Big),\\
\delta b_t &=& \delta b^{(0)}_t (x_i)+ \cO(r), \qquad \delta b_i = \frac{\ln r }{4 \pi T} \Big( - B_i + N_d \xi_i \Big) + \cO(r),\\
\delta b_r &=& \frac{1}{U(r)} \Big( \delta b^{(0)}_r(x_i) + \cO(r) \Big),
\label{hor-pert}
\een
It turns out that the constraint on the leading order have to be imposed
\ben
\delta g^{(0)}_{tt} &+& \delta g^{(0)}_{rr} - 2 \delta g^{(0)}_{rt} = 0, \qquad \delta g^{(0)}_{ri} = \delta g^{(0)}_{ti},\\ \nonumber
\delta a^{(0)}_r &=& \delta a^{(0)}_t, \qquad \delta b^{(0)}_r = \delta b^{(0)}_t.
\een

\subsection{ Equations for perturbations at the event horizon }
One imposes on a subset of the linearized black brane perturbations, i.e., $\delta g^{(0)}_{it}, ~\delta g^{(0)}_{rt},~\delta a^{(0)}_t, ~\delta b^{(0)}_t,$
the relations as follows \cite{rog19}:
\ben \label{F}
\na_i \na^i w &+& \na_i E^i + \na_i \Big( a^{(0)}_t v^i \Big) + \frac{\alpha}{2} \Big[ \na_m \na^m w_d + \na_m B^m + \na_m \Big( b^{(0)}_t v^m \Big) \Big] = 0,\\ \label{B}
\na_i \na^i w_d &+& \na_i B^i + \na_i \Big( b^{(0)}_t v^i \Big) + \frac{\alpha}{2} \Big[ \na_m \na^m w + \na_m E^m + \na_m \Big( a^{(0)}_t v^m \Big) \Big] = 0,\\ \label{stokes}
b^{(0)}_t \Big[ \na_i w_d &+& B_i + \frac{\alpha}{2} \Big( \na_i w + E_i \Big) \Big]
+ a^{(0)}_t \Big[ \na_i w + E_i + \frac{\alpha}{2} \Big( \na_i w_d + B_i \Big) \Big] \\ \nonumber
&-& \na_i \phi^{(0)} \na_m \phi^{(0)} ~v^m + 2~\na^m \na_{(m} v_{i)} + 4 \pi T \xi_i - \na_i p = 0,\\ \label{vv}
\na_i v^i &=& 0,
\een
where we have denoted
\ben \label{pp1}
w &=& \delta a^{(0)}_t, \qquad w_d = \delta b^{(0)}_t, \qquad p = - 4 \pi T \frac{\delta g^{(0)}_{rt}}{G^{(0)}} - \delta g^{(0)}_{it} ~ \na^i \ln G^{(0)},\\ \nonumber
v_i &=& - \delta g^{(0)}_{it}.
\een
The above equations  result from the conservation of charge and heat currents 
in the unperturbed system.
The  variables introduced in (\ref{pp1}) are a subset of all perturbations.
They are found to fulfill at the horizon the generalized Navier-Stokes equations (\ref{F}) - (\ref{vv}). 
As discussed earlier for the single current model \cite{ban15} the scalar field contributes viscosity
like term as also does the curvature of the horizon. The latter is best visible by writing 
\be
2~\na^m \na_{(m} v_{j)}=\na^2v_j+R_{ji}v^i.
\ee

\section{Conductivity for the uncharged black brane}\label{sec:cond}
In this section we shall consider the case without $\phi$ field, responsible for dissipation. 
Further, for the connectedness with \cite{gro15} we define the quantities
\be
Q= a^{(0)}_t, \qquad Q_d = b^{(0)}_t, \qquad \na_j w = - \na_j \mu, \qquad \na_j w_d = - \na_j \mu_d.
\ee
Let us first study the conductivities for Dirac semimetals in the uncharged black object case,
 i.e., $Q =0,~Q_d =0$. For the considered situation equation (\ref{F}) and (\ref{B}) decoupled to the relations
\ben \label{ff}
\na_i \Big[ \sqrt{g^{(0)}} \Big( E^i &-& \na^i \mu \Big) \Big] = 0,\\ \label{bb}
\na_i \Big[ \sqrt{g^{(0)}} \Big( B^i &-& \na^i \mu_d \Big) \Big] = 0.
\een
Because of the fact that they constitute the linear equations, we may set 
\be
\mu = \mu_a ~E^a, \qquad \mu_d = {\mu_d}_a ~B^a,
\ee
in the equations (\ref{ff}) and (\ref{bb}). This substitution reveals that
\ben
\na_i \Big[ \sqrt{g^{(0)}} \Big( \delta^i{}_k &-& \na^i \mu_k \Big) E^k \Big] = 0,\\
\na_i \Big[ \sqrt{g^{(0)}} \Big( \delta^i{}_k &-& \na^i {\mu_d}_k \Big) B^k\Big] = 0.
\een
On the other hand,  it follows that for some constants $\psi^i{}_k,~{(\psi_d})^i_k$ and functions $\Psi^i,~{\Psi_d}^i$, one obtains
\ben \label{c1}
\sqrt{g^{(0)}} g^{ij}_{(0)}  \Big( \delta^k{}_j &-& \na_j \mu^k \Big) = \ep^{ij} \Big( \psi^i{}_k - \na_j \Psi^k \Big),\\ \label{c2}
\sqrt{g^{(0)}} g^{ij}_{(0)} \Big( \delta^k{}_j &-& \na_j \mu^k_d \Big) = \ep^{ij} \Big( ({\psi_d})^i_k - \na_j \Psi_d^k \Big),
\een
Using the properties of the antisymmetric two-dimensional tensor $\ep^{ij}$, it can be proved, that the above equations are equivalent to
\ben \label{1f}
- \ep^{am}\Big( \delta^k{}_m &-& \na_m \mu^k \Big) = \sqrt{g^{(0)}} g^{aj}_{(0)} \Big( \psi^k {}_j - \na_j \Psi^k \Big),\\ \label{1d}
- \ep^{am} \Big( \delta^k{}_m &-& \na_m \mu_d^k \Big) = \sqrt{g^{(0)}} g^{aj}_{(0)} \Big( (\psi_d)^k {}_j - \na_j \Psi_d^k \Big),
\een
Taking the spatial derivatives of the relations (\ref{1f}) and (\ref{1d}), using the uniqueness and linearity arguments, one obtains that
\ben \label{2f}
\na_i \Psi^k &=&  \psi^k {}_j ~\na_i \mu^j,\\ \label{2d}
\na_i \Psi_d^k &=&  (\psi_d)^k {}_j ~\na_i \mu_d^j.
\een
Combining relations (\ref{1f})-(\ref{1d}) and (\ref{2f})-(\ref{2d}), we arrive at the following:
\ben \label{3f}
\sqrt{g^{(0)}} g^{ij}_{(0)}  \Big( \delta^k{}_j &-& \na_j \mu^k \Big) = - \ep^{ij} \Big( \delta^r{}_j - \na_j \mu^r \Big) (\psi^{-1})^k_r,\\ \label{3d}
\sqrt{g^{(0)}} g^{ij}_{(0)}  \Big( \delta^k{}_j &-& \na_j {\mu_d}^k \Big) = - \ep^{ij} \Big( \delta^r{}_j - \na_j \mu^r \Big) (\psi_d^{-1})^k_r,
\een
Multiplying equation (\ref{3f}) by $E_k$ and relation (\ref{3d}) by $B_k$, we arrive at the component of the gauge currents $J^i_{(F)},~J^i_{(B)}$, respectively.
Having in mind that neglecting heat transport the gauge currents can be written as
\ben \label{curf}
J^i_{(F)} &=& \sigma^{ij}_{FF} E_j + \sigma^{ij}_{FB} B_j,\\ \label{curb}
J^i_{(B)} &=& \sigma^{ij}_{BF} E_j + \sigma^{ij}_{BB} B_j,
\een
we arrive at the conclusion that the DC conductivities constitute relations as follows:
\ben \label{ss}
\sigma^{ij}_{FF}  &=& \ep^{im} ~\psi^j{}_m, \qquad
\sigma^{ij}_{FB} = \frac{\alpha}{2} \ep^{im} ~\psi^j{}_m, \\ \label{ss1}
\sigma^{ij}_{BB}  &=& \ep^{im} ~(\psi_d)^j{}_m, \qquad
\sigma^{ij}_{BF} = \frac{\alpha}{2} \ep^{im} ~(\psi_d)^j{}_m.
\een
The inspection of the equations (\ref{c1})-(\ref{c2}) and (\ref{3f})-(\ref{3d}), we draw a conclusion that their consistency is ensured if
$\psi_a{}^k ~\psi_k{}^b = - \delta_a{}^b$ and similarly $(\psi_d)_{a}{}^k ~(\psi_d)_{k}{}^b = - \delta_a{}^b$. On the other hand, the relations (\ref{ss}) and (\ref{ss1})
enable us to write
\ben
\det \sigma_{FF} &=& \frac{1}{2!} \ep^{ab}~\ep_{mk} ~(\sigma_{FF})_{a}{}^{m} (\sigma_{FF})_b{}^k = \frac{1}{2!} \ep^{ab}~\ep_{mk} ~\psi_{a}{}^{m} \psi_b{}^k, \\
\det \sigma_{BB} &=& \frac{1}{2!} \ep^{ab}~\ep_{mk} ~(\sigma_{BB})_{a}{}^{m} (\sigma_{BB})_b{}^k = \frac{1}{2!} \ep^{ab}~\ep_{mk} ~(\psi_d)_{a}{}^{m} (\psi_d)_b{}^k, \\
\det \sigma_{FB} &=& \frac{1}{2!} \ep^{ab}~\ep_{mk} ~(\sigma_{FF})_{a}{}^{m} (\sigma_{FF})_b{}^k = \frac{1}{2!} \frac{\alpha^2}{4} \ep^{ab}~\ep_{mk}~ \psi_{a}{}^{m} \psi_b{}^k, \\
\det \sigma_{BF} &=& \frac{1}{2!} \ep^{ab}~\ep_{mk} ~(\sigma_{BB})_{a}{}^{m} (\sigma_{BB})_b{}^k = \frac{1}{2!}\frac{\alpha^2}{4}  \ep^{ab}~\ep_{mk} 
~(\psi_d)_{a}{}^{m} (\psi_d)_b{}^k.
\een
Having in mind the consistency condition, mentioned above, one arrives at the following:
\be
\det \sigma_{FF} = 1, \qquad \det \sigma_{BB} = 1,
\label{bound-d}
\ee
which implies that
\be 
\det \sigma_{FB} = \frac{\alpha^2}{4}, \qquad \det \sigma_{BF} = \frac{\alpha^2}{4}.
\label{bound-nd}
\ee

Consequently, for the determinant of the conductivity in the theory under consideration, we obtain
\be
\det \sigma = \Big( \frac{1}{2!} \Big)^2~\beta~~\ep_{b j_1}~ \ep_{mk}~\ep^{d j_2}~\ep_{sz}
\psi^m{}_b~\psi^k{}_{j_1}~(\psi_d)^s{}_d~(\psi_d)^z{}_{j_2} = \beta ~\det \sigma_{FF}~\det \sigma_{BB} = \beta,
\label{bound1}
\ee
where  we set
\be
\beta =\talpha~\Big( 1 + \frac{\alpha^2}{4} \Big),
\label{betab1}
\ee
and $\talpha = 1 - \frac{\alpha^2}{4}$.

Let us assume that the conductivities $\sigma_{ab},~a,b=F,~B$ do not depend on the spatial directions. Under these circumstances they can be considered as scalars. However, the full conductivity $\sigma$ of the system is the $2\times 2$ matrix with entries \{$\sigma_{FF},\sigma_{FB};\sigma_{BF},\sigma_{BB}$\} and thus on the basis of 
the equations (\ref{bound-d}) and (\ref{bound-nd}) one has that $det\sigma=\sigma_{FF}\sigma_{BB}-\sigma_{FB}\sigma_{BF}=1-(\alpha/2)^4=\beta$.  
It can be seen that
the considered bound is also valid for $\alpha=0$, when the matrix is diagonal $\sigma=diag\{\sigma_{FF},\sigma_{BB}\}$.

\section{Charged black brane case}
\label{sec:charged}
In order to study the charged case, let us define for the adequate $U(1)$-gauge field, bulk dual tensors \cite{her07}
\be
\mathbb{F}^{rj} = - \frac{1}{2} \frac{\ep^{rjab}}{\sqrt{-g}} F_{ab}, \qquad
\mathbb{B}^{rj} = - \frac{1}{2} \frac{\ep^{rjab}}{\sqrt{-g}} B_{ab},
\ee
with the property $\ep^{rtxy} = \frac{1}{\sqrt{-g}}$, as $r \rightarrow \infty$.
It implies that one can find constant dual currents densities connected with Maxwell and auxiliary gauge fields, in the boundary theory, which imply
\be
I^i_{(F)} = \ep^{ij} \Big( E_j + \frac{\alpha}{2} B_j \Big), \qquad I^i_{(B)} = \ep^{ij} \Big( B_j + \frac{\alpha}{2} E_j \Big).
\ee
The equations of motion $\p_i J^i_{(F)} = 0,~\p_i J^i_{(B)} = 0$ yield that
\ben
\p_r \mathbb{E}\Big[ \mathbb{F}^{ir} &+& \frac{\alpha}{2} \mathbb{B}^{ir} \Big] =0,\\
\p_r \mathbb{E}\Big[ \mathbb{B}^{ir} &+& \frac{\alpha}{2} \mathbb{F}^{ir} \Big] =0,
\een
where we have denoted the spatial average by $ \mathbb{E}[M]=\frac{1}{L^2} \int dx^2 M$, for the coordinates which 
satisfy the periodic boundary conditions
$x_i \rightarrow x_i + L$.

Just using the duals, at $r \rightarrow 0$, we get
\ben
\mathbb{E}\Big[ \mathbb{F}^{it} &+& \frac{\alpha}{2} \mathbb{B}^{it} \Big] = \ep^{im} J_{(F) m},\\
\mathbb{E}\Big[ \mathbb{B}^{it} &+& \frac{\alpha}{2} \mathbb{F}^{it} \Big] = \ep^{im} J_{(B) m}.
\een

The spatial averaged dual electric currents connected with Maxwell and auxiliary fields, are independent on the radius of the bulk. It means that they can be defined
on the black object event horizon
\ben
\mathbb{E}\Big[ \mathbb{F}^{it} &+& \frac{\alpha}{2} \mathbb{B}^{it} \Big] = 
\cE_j + \frac{\alpha}{2} \cB_j = \tau_{(F)}^{j k}~{I_{(F)}}_k , \\
\mathbb{E}\Big[ \mathbb{B}^{it} &+& \frac{\alpha}{2} \mathbb{F}^{it} \Big] = 
\cB_j + \frac{\alpha}{2} \cE_j = \tau_{(B)}^{jk}~{I_{(B)}}_k,
\een
where $\tau_{(F)}^{j k}$ and $\tau_{(B)}^{j k}$ are the dual resistivity tensors bounded with Maxwell and {\it hidden sector} gauge fields.
One can relate $\cE_i,~\cB_i,~I^i_{(F)},~I^i_{(B)},~J_{(F)}^k,~J_{(B)}^k$ and obtain
\ben
J_{(F) a} &=& \ep_{ia}~\ep_{jk}~\tau_{(F)}^{ij} \Big( E^k + \frac{\alpha}{2} B^k \Big), \\
J_{(B) a} &=& \ep_{ia}~\ep_{jk}~\tau_{(B)}^{ij} \Big( B^k + \frac{\alpha}{2} E^k \Big).
\een
It leads to the following relations:
\ben
\sigma_{FF}^{ij} &=& \ep^{mi}~\ep^{n j}~\tau_{mn}^{(F)},\\
\sigma_{FB}^{ij} &=& \ep^{mi}~\ep^{n j}~\tau_{mn}^{(F)}~\frac{\alpha}{2},\\
\sigma_{BB}^{ij} &=& \ep^{mi}~\ep^{n j}~\tau_{mn}^{(B)},\\
\sigma_{BF}^{ij} &=& \ep^{mi}~\ep^{n j}~\tau_{mn}^{(B)}~\frac{\alpha}{2}.
\een
Consequently, it can be found that
\ben
\det \sigma_{FF}&=& \frac{1}{2!}~\ep^{mi}~\ep^{n j}~\tau_{mn}^{(F)},\\
\det \sigma_{FB}&=& \frac{1}{2!} \frac{\alpha^2}{4}\ep^{mi}~\ep^{n j}~\tau_{mn}^{(F)},\\
\det \sigma_{BB}&=& \frac{1}{2!}~\ep^{mi}~\ep^{n j}~\tau_{mn}^{(B)},\\
\det \sigma_{BF}&=& \frac{1}{2!} \frac{\alpha^2}{4}\ep^{mi}~\ep^{n j}~\tau_{mn}^{(B)}.
\een
As in \cite{gro15} we assume that the boundary theory constitutes the particle vortex dual, which leads to the conjecture that
\be
\det \sigma_{FF} = \frac{1}{\det \tau_{(F)}}, \qquad \det \sigma_{BB} = \frac{1}{\det \tau_{(B)}},
\ee
which in turn precedes to the conditions
\be
\Big( \det \tau_{(F)} \Big)^2 =1, \qquad \Big( \det \tau_{(B)} \Big)^2 =1.
\ee
By virtue of the above relations, in the charge case the determinant of the conductivity is given by
\be
\det \sigma = \Big( \frac{1}{2!} \Big)^2~\beta_1~\ep_{i_1 k}~ \ep_{j_1 l}~\ep_{i_2 a}~\ep_{j_2 b}
\tau^{i_1 j_1}_{(F)}~\tau^{k l}_{(F)}~\tau^{i_2 j_2}_{(B)}~\tau^{ab}_{(B)} = \beta_1~\det \tau_{(F)}~\det \tau_{(B)} = \beta_1,
\label{bound2}
\ee
where $\beta_1$ is given as follows:
\be
\beta_1 = \talpha~\Big( 1 + \frac{\alpha^2}{4} \Big).
\ee
The bound we have obtained in the charged case, is the same  as in the uncharged model
found earlier. The parameter $\beta_1$ is monotonously diminishing function of the
coupling $\alpha$ from its canonical value 1, when $\alpha \rightarrow 0$ and to the value 0
for $\alpha \rightarrow 2$. 
On physical grounds one expects $\alpha\le 1$. The theory predicts the
lowering of the bound from its the value 1 towards $\approx 0.94$ at $\alpha =1$.

\section{Variational attitude}
\label{sec:variation}
This section will be devoted to the variational techniques implemented in order to establish 
the lower bounds on DC-conductivities. The bounds will be achieved in an analogous way as the upper bounds of resistance of a disordered 
resistor network, based on the Thomson's principle \cite{lev09}-\cite{luc15}. It states that if one runs a set 
of 'trial' currents through a resistor network, being subject to 
certain boundary conditions, the upper bound of the inverse conductivity can be computed by applying the variational
principle to the power dissipated by the 'trial' currents in question. 
It happens that the power dissipated by 'trial' currents is minimal for the true distribution 
of the aforementioned currents.

To proceed further, let us recall that the Stokes equation on the black brane event horizon can be recast in the form as derived in Ref. \cite{rog19}
\ben \nonumber \label{pos}
\int \sqrt{g^{(0)}} d^2 x \Big[ 
 2 \na^{(i}v^{j)} \na_{(i}v_{j)} &+& \Big( \na_i w + E_i \Big)\Big( \na^i w + E^i \Big) + \Big( \na_i w_d + B_i \Big)\Big( \na^i w_d + B^i \Big)\\
 \alpha \Big( \na_i w &+& E_i \Big)\Big( \na^i w_d + B^i \Big)
+ v^m \na_m \phi^{(0)} \na_j \phi^{(0)} v^j \Big] \\ \nonumber
&=& \int d^2x \Big[ Q^{i (0)}  \xi_i + J^ {i (0)}_{(F)}  E_i + J^ {i (0)}_{(B)}  B_i \Big],
\een
with the the adequate definitions of the currents, given by
\ben
J^{i (0)}_{(F)} = J^I_{(F)} \mid_{\cH} &=& \sqrt{g^{(0)}} g^{ij}_{(0)}~
\Big[
\Big( \na_j (\delta a^{(0)}_t) + E_j - a^{(0)}_t ~\delta g^{(0)}_{tj} \Big) \\ \nonumber
&+& \frac{\alpha}{2} \Big( \na_j (\delta b^{(0)}_t) + B_j - b^{(0)}_t ~\delta g^{(0)}_{tj} \Big) \Big],\\
J^{i (0)}_{(B)} = J^I_{(B)} \mid_{\cH} &=& \sqrt{g^{(0)}} g^{ij}_{(0)}~
\Big[
\Big( \na_j (\delta b^{(0)}_t) + B_j - b^{(0)}_t ~\delta g^{(0)}_{tj} \Big) \\ \nonumber
&+& \frac{\alpha}{2} \Big( \na_j (\delta a^{(0)}_t) + E_j - a^{(0)}_t ~\delta g^{(0)}_{tj} \Big) \Big],\\
Q^{i (0)} = Q^{i} \mid_{\cH} &=& - 4 \pi~ T ~\sqrt{g^{(0)}} g^{ij}_{(0)}~\delta g^{(0)}_{tj}.
\een
Additionally one has that the following conservation relations are fulfilled:
\be
\na_i J^{i (0)}_{(F)} = 0, \qquad \na_i J^{i (0)}_{(B)} = 0, \qquad \na_i Q^{i (0)} = 0.
\ee
Further, let us define
\be
J^{i (0)}_{(F)} = \sqrt{g^{(0)}}  \cJ^i_F, \qquad J^{i (0)}_{(F)} = \sqrt{g^{(0)}}  \cJ^i_F, \qquad Q^{i (0)} = 4 \pi T \sqrt{g^{(0)}}  v^i.
\ee
Consequently the relation (\ref{pos}) may be rewritten in the form
\ben \nonumber \label{sto}
\int d^2x \Big[ Q^{i (0)}  \xi_i &+& J^ {i (0)}_{(F)}  E_i + J^ {i (0)}_{(B)}  B_i \Big] =
\int \sqrt{g^{(0)}} d^2 x \Bigg[ 
 2 \na^{(i}v^{j)} \na_{(i}v_{j)} + v^m \na_m \phi^{(0)} \na_j \phi^{(0)} v^j \\
 + ~\Big[ \frac{1}{\talpha} \Big( \cJ^i_F &-& \frac{\alpha}{2} \cJ^i_B \Big) - a^{(0)}_t v^i\Big]
\Big[ \frac{1}{\talpha} \Big( \cJ_{i F} - \frac{\alpha}{2} \cJ_{i B} \Big) - a^{(0)}_t  v_i\Big] \\ \nonumber
+ ~\Big[ \frac{1}{\talpha} \Big( \cJ^i_B &-& \frac{\alpha}{2} \cJ^i_F \Big) - b^{(0)}_t v^i\Big]
\Big[ \frac{1}{\talpha} \Big( \cJ_{i B} - \frac{\alpha}{2} \cJ_{i F} \Big) - b^{(0)}_t  v_i\Big] \\ \nonumber
+ ~\alpha ~\Big[ \frac{1}{\talpha} \Big( \cJ^i_F &-& \frac{\alpha}{2} \cJ^i_B \Big) - a^{(0)}_t v^i\Big]
\Big[ \frac{1}{\talpha} \Big( \cJ_{i B} - \frac{\alpha}{2} \cJ_{i F} \Big) - b^{(0)}_t  v_i\Big] \Bigg],
\een
which is the subject of the following analysis.

\subsection{Bound on conductivities}
In order to establish the bounds on the conductivities in the holographic model 
of graphene, we shall analyze the left-hand side of the equation (\ref{sto}),
which includes the definition of the dissipated power.
The dissipated power will be provided by the following expression:
\be
P = J^i_F~E_i + J^i_B~B_i + Q^i \xi_i,
\label{pow}
\ee
where the above quantities are normalized by averaging them spatially over the black brane event horizon, i.e.,
\be
J^i_F = \mathbb{E} [ J^{i (0)}_{(F)} ], \qquad J^i_B = \mathbb{E} [ J^{i (0)}_{(B)} ], \qquad Q^i =  \mathbb{E} [ Q^{i (0)} ].
\ee
In what follows, we consider compact and flat spatial dimensions of the dual theory. 

Using equation (\ref{sto}) the dissipative power (\ref{pow}) implies 
\ben \nonumber \label{pp}
P &=& \mathbb{E}\Bigg[ 
 2 \na^{(i}v^{j)} \na_{(i}v_{j)} + v^m \na_m \phi^{(0)} \na_j \phi^{(0)} v^j \\ \nonumber
 &+& ~\Big[ \frac{1}{\talpha} \Big( \cJ^i_F - \frac{\alpha}{2} \cJ^i_B \Big) - a^{(0)}_t v^i\Big]
\Big[ \frac{1}{\talpha} \Big( \cJ_{i F} - \frac{\alpha}{2} \cJ_{i B} \Big) - a^{(0)}_t  v_i\Big] \\
&+& ~\Big[ \frac{1}{\talpha} \Big( \cJ^i_B - \frac{\alpha}{2} \cJ^i_F \Big) - b^{(0)}_t v^i\Big]
\Big[ \frac{1}{\talpha} \Big( \cJ_{i B} - \frac{\alpha}{2} \cJ_{i F} \Big) - b^{(0)}_t  v_i\Big] \\ \nonumber
&+& ~\alpha ~\Big[ \frac{1}{\talpha} \Big( \cJ^i_F - \frac{\alpha}{2} \cJ^i_B \Big) - a^{(0)}_t v^i\Big]
\Big[ \frac{1}{\talpha} \Big( \cJ_{i B} - \frac{\alpha}{2} \cJ_{i F} \Big) - b^{(0)}_t  v_i\Big] \Bigg].
\een
One can consider $P$ as a functional of $v^i$ and the $U(1)$-gauge currents. It means that for arbitrary conserved periodic set of charge and heat currents directed along
$v^i$ one has that
\be
\cJ^i_F = {\tilde \cJ}^i_F + {\tilde {\tilde \cJ}}{}^i_F, \qquad \cJ^i_B = {\tilde \cJ}^i_B + {\tilde {\tilde \cJ}}{}^i_B,
\qquad v^i = {\tilde v}^i + {\tilde {\tilde v}}{}^i,
\ee
where $( {\tilde v}^i,~{\tilde \cJ}{}^i_F,~{\tilde \cJ}{}^i_B)$ stands for the exact solution of the underlying system of hydrodynamical equations, being
subject to the adequate boundary conditions. On the other hand,
$( {\tilde {\tilde v}}{}^i,~ {\tilde {\tilde \cJ}}{}^i_F,~ {\tilde {\tilde \cJ}}{}^i_B)$ denote the deviations from the exact solution. Expansion of 
$P$ reveals that we get
\ben
P[v^i,~\cJ^i_F ,~\cJ^i_B] &=& P[ {\tilde v}^i + {\tilde {\tilde v}}{}^i,~{\tilde \cJ}^i_F + {\tilde {\tilde \cJ}}{}^i_F,~{\tilde \cJ}^i_B + {\tilde {\tilde \cJ}}{}^i_B]\\ \nonumber
&=& P[{\tilde v}^i,~{\tilde \cJ}^i_F,~{\tilde \cJ}^i_B] + P[{\tilde {\tilde v}}{}^i,~{\tilde {\tilde \cJ}}{}^i_F,~{\tilde {\tilde \cJ}}{}^i_B] + 2 K,
\een
where the quantity $2 K$, implies
\ben
2K &=& 2 \Big[ \frac{1}{\talpha} \Big( 
{\tilde \cJ}^i_F - \frac{\alpha}{2} \cJ^i_B \Big) - a^{(0)}_t {\tilde v}^i\Big]
\Big[ \frac{1}{\talpha} \Big( {\tilde {\tilde {\cJ}}}_{i F} - \frac{\alpha}{2} {\tilde {\tilde {\cJ}}}_{i B} \Big) - a^{(0)}_t  {\tilde {\tilde v}}_i\Big] \\ \nonumber
&+&
2 \Big[ \frac{1}{\talpha} \Big( 
{\tilde \cJ}^i_B - \frac{\alpha}{2} \cJ^i_F \Big) - a^{(0)}_t {\tilde v}^i\Big]
\Big[ \frac{1}{\talpha} \Big( {\tilde {\tilde {\cJ}}}_{i B} - \frac{\alpha}{2} {\tilde {\tilde \cJ}}_{i F} \Big) - a^{(0)}_t  {\tilde {\tilde v}}_i\Big] \\ \nonumber
&+&
 \Big[ \frac{1}{\talpha} \Big( 
{\tilde \cJ}^i_F - \frac{\alpha}{2} \cJ^i_B \Big) - a^{(0)}_t {\tilde v}^i\Big]
\Big[ \frac{1}{\talpha} \Big( {\tilde {\tilde {\cJ}}}_{i B} - \frac{\alpha}{2} {\tilde {\tilde \cJ}}_{i F} \Big) - a^{(0)}_t  {\tilde {\tilde v}}_i \Big] \\ \nonumber
&+&
 \Big[ \frac{1}{\talpha} \Big( 
{\tilde \cJ}^i_B - \frac{\alpha}{2} \cJ^i_F \Big) - a^{(0)}_t {\tilde v}^i\Big]
\Big[ \frac{1}{\talpha} \Big( {\tilde {\tilde {\cJ}}}_{i B} - \frac{\alpha}{2} {\tilde {\tilde {\cJ}}}_{i F} \Big) - a^{(0)}_t  {\tilde {\tilde v}}_i\Big] \\ \nonumber
&+&
4 \na^{(i}{\tilde v}^{j)} \na_{(i} {\tilde {\tilde v}}_{j)} + 2 {\tilde {\tilde v}}^m \na_m \phi^{(0)} \na_i \phi^{(0)} {\tilde v}^i.
\een
It can be shown using the current equations and integration by parts that $K=0$. Consequently, it reveals that
\be
P[v^i,~\cJ^i_F ,~\cJ^i_B] \ge P[{\tilde v}^i,~{\tilde \cJ}^i_F,~{\tilde \cJ}^i_B] .
\ee
As was explained in \cite{gro15}, for the charged black brane one may set $v_i =0$, which trivially fulfils the constraints equations.
Then we arrive at 
\be
P[0,~\cJ^i_F ,~\cJ^i_B] = \int \sqrt{g^{(0)}} d^2 x ~\frac{\ep^{i_1}{}_a~\ep^{j_1}{}_b }{\det \sigma}
\Bigg[ 
\sigma^{BB}_{i_1 j_1} ~\cJ^a_{F} \cJ^b_F + \sigma^{FF}_{i_1 j_1} ~\cJ^a_{B} \cJ^b_B - \Big(  \sigma^{FB}_{i_1 j_1} + \sigma^{BF}_{i_1 j_1} \Big)
~\cJ^a_{F} \cJ^b_B \Bigg].
\label{reduk0}
\ee 
On the other hand, using the equation (\ref{pp}), one arrives at the following:
\be \label{reduk1}
P[0,~\cJ^i_F ,~\cJ^i_B] = \int \sqrt{g^{(0)}} d^2 x \Bigg[ 
\frac{1}{\talpha} \Big( \cJ^i_F \Big)^2 + \frac{1}{\talpha} \Big( \cJ^i_B \Big)^2 - \frac{\alpha}{\talpha} \cJ^i_F \cJ_{i B} 
\Bigg].
\ee
Comparison of the equations (\ref{reduk0}) and (\ref{reduk1}) give us the conditions imposed on the electrical conductivities, in general case.

To commence with, let us analyze limits of the obtained relations. 
Firstly one supposes that in the absence of the heat current, 
we shall consider only the single current case.
In this case $B_i =0,~ \alpha =0$ and the relations (\ref{curf})-(\ref{curb}) reveal that 
\be
\sigma^{ij}_{BF} =0, \qquad \sigma^{ij}_{FB} =0, \qquad \sigma^{ij}_{BB} = 0,
\ee
and $E^i = \frac{\cJ^i_F}{\sigma_{FF}}$. Taking into account (\ref{curf}) and calculating the dissipative power we get
\be
P = \int \sqrt{g^{(0)}} d^2x ~\frac{\cJ^i_F~\cJ_{i F}}{\sigma_{FF}} = \int \sqrt{g^{(0)}} d^2 x~\cJ^i_F~\cJ_{i F}.
\ee
It implies that the following relation takes place:
\be
\sigma_{FF} \ge 1.
\ee
Consequently, for the model with only auxiliary $ U(1)$-gauge field,
one has that 
\be
E_i=0, \qquad \sigma^{ij}_{FB} =0, 
\ee
and $B_i =\frac{\cJ^i_B}{\sigma_{BB}}$. 
The same  reasoning as above leads to the relation
\be
P = \int \sqrt{g^{(0)}} d^2x ~\frac{\cJ^i_B~\cJ_{i B}}{\sigma_{BB}} = \frac{1}{\talpha}~\int \sqrt{g^{(0)}} d^2 x~\cJ^i_B~\cJ_{i B},
\ee
and it yields that
\be
\sigma_{BB} \ge \talpha.
\ee

In the next step, because of the complexity of the exact relations, let us suppose 
that the existence only of  $\cJ^x_F$ and $\cJ^x_B$ currents.
By straightforward calculations it can be envisaged that $P[0,~\cJ^i_F ,~\cJ^i_B] $ reduces to
\be \label{reduk}
P[0,~\cJ^i_F ,~\cJ^i_B] = \int \sqrt{g^{(0)}} d^2 x ~\frac{1}{\det \sigma} \Bigg[ 
\sigma^{BB}_{yy} \Big( \cJ^x_F \Big)^2 + \sigma^{FF}_{yy} \Big( \cJ^x_B \Big)^2 
- \Big( \sigma^{FB}_{yy} + \sigma^{BF}_{yy} \Big) \cJ^x_F \cJ^x_B 
\Bigg].
\ee
Comparing the relations (\ref{reduk}) and (\ref{reduk1}), the estimations for the adequate components of $\sigma^{ij}_{\alpha \beta}$ tensor can be achieved
\be
\frac{\sigma^{BB}_{yy}}{\det \sigma} = \frac{1}{\talpha}, \qquad \frac{\sigma^{FF}_{yy}}{\det \sigma} = \frac{1}{\talpha},
\qquad \frac{\sigma^{FB}_{yy} + \sigma^{BF}_{yy}}{\det \sigma} = \frac{\alpha}{\talpha}.
\ee

\section{Summary and conclusions}
\label{sum-concl}
In our paper we have studied the lower bounds 
of the electrical conductivities in the holographic model 
of the strongly interacting two-dimensional graphene sheet with disorder by means of the gauge-gravity duality. 
It happens that graphene close to the particle - hole symmetry point is a laboratory 
system fulfilling the {strong coupling  requirements.
 On the gravity side we elaborate the Einstein-Maxwell theory supplemented by the auxiliary 
$U(1)$-gauge field. The ordinary Maxwell and the auxiliary fields are coupled by the
{\it kinetic mixing} term, with a coupling constant $\alpha$. In the studies we pay attention 
to the linear  response of the black brane to the electric fields
of the aforementioned gauge fields. On the field theory side, the situation
coincides with the existence of two  transport currents, which in graphene may correspond to electron and hole currents.
The mixing parameter $\alpha$ may be responsible for the phase space constraints of scattering events in
system with Dirac spectrum.

We have found the modifications of the bounds due to the coupling between 
the currents. For the physically expected values of $\alpha$-coupling constant 
which is smaller than 1, the obtained bound $\beta$
for the conductivity tensor $\sigma$, $\det\sigma= \beta$ is only slightly below 1. 

It would be of interest to analyze the 
existence of the similar bound in Dirac or Weyl semimetals, being the three-dimensional analogues of graphene.

\acknowledgments
KIW was partially supported by the grant DEC-2017/27/B/ST3/01911 of the National Science Center (Poland).





\begin{thebibliography}{99}
\bibitem{rey}
The Reynolds number $Re$ is the ratio of the inertial 
to viscous forces and can be expressed as $Re=\frac{\rho u L}{\mu}$, where $\rho$ is fluid density,
$u$ the velocity, $L$ the characteristic length affecting the flow and $\mu$ [Pa$\cdot$ s] 
the dynamic viscosity.
\end{thebibliography}
\end{document}